\documentclass{article}
\usepackage[latin1]{inputenc}
\usepackage{graphicx}
\usepackage{fancyhdr}
\usepackage{amsfonts}

\usepackage{apalike}
\author{
Ueli Rutishauser\\
Computation and Neural Systems, California Institute of Technology\\
\texttt{urut@caltech.edu} \\
\and
Rodney J. Douglas \\
Institute of Neuroinformatics, ETH and University of Zurich \\
\texttt{rjd@ini.phys.ethz.ch} \\
}

\title{State dependent computation using coupled recurrent networks
\thanks{The final version of this article will be published in Neural Computation, published by The MIT Press. See http://mitpress.mit.edu/NECO . The final article is already available as
\emph{early access} from MIT Press.}
}

\begin{document}
\maketitle

\begin{abstract}
\begin{normalsize}
Although conditional branching between possible behavioural states is a hallmark of intelligent behavior, very little is known about the neuronal mechanisms that support this processing. In a step toward solving this problem we demonstrate by theoretical analysis and simulation how networks of richly inter-connected neurons, such as those observed in the superficial layers of the neocortex, can embed reliable robust finite state machines. We show how a multi-stable neuronal network containing a number of states can be created very simply, by coupling two recurrent networks whose synaptic weights have been configured for soft winner-take-all (sWTA) performance. These two sWTAs have simple, homogenous locally recurrent connectivity except for a small fraction of recurrent cross-connections between them, which are used to embed the required states. This coupling between the maps allows the network to continue to express the current state even after the input that elicted that state is withdrawn. In addition, a small number of 'transition neurons' implement the necessary input-driven transitions between the embedded states. We provide simple rules to systematically design and construct neuronal state machines of this kind. The significance of our finding is that it offers a method whereby the cortex could construct networks supporting a broad range of sophisticated processing by applying only small specializations to the same generic neuronal circuit.
\end{normalsize}

\end{abstract}

\section{Introduction}
Quantitative studies of the anatomical connection weights between neurons in cat visual cortex have revealed that one prominent feature of the neocortical circuit of cat visual cortex is the high degree of connectivity between pyramidal cells in the superficial cortical layers \cite{Binzegger04}. About 30\% of all excitatory synapses onto these cells are derived from other superficial pyramids, and most of these connections are short-range (arising from source neurons within about $300~/mu$). In addition to excitatory inputs, these pyramids also receive inhibitory inputs, which consititute about 10\% of their synaptic input \cite{Shepherd05,Binzegger04}. Thus, the activation of these excitatory neurons can be strongly affected by both positive and negative feedback loops. In theory, these recurrent circuits exhibit a variety of interesting computations \cite{Abbott94,Douglas95,BenYishai95,Hahnloser00,Li01,Pouget98,Machens05,Wang02,Rao04,Hochreiter97,Coultrip92}. For example, the soft winner-take-all (sWTA) network is able to selectively enhance one part of its input while suppressing the remainder \cite{Hahnloser99,Maass00}, and so offers a form of signal restoration between computational stages sought after by von Neumann in his early explorations of brain-like principles of computation \cite{vonNeumann58}. 

In theoretical models, the neurons that compose a WTA are usually organized as a simple linear map in which each unit receives excitatory inputs from its neighbouring units as well as an inhibitory input that is proportional to the total activity of all units \cite{AbbottDayan01,Douglas2007_recurrent}. We use the term 'map' to indicate that the elements of the WTA network are not functionally independent. Rather, their activity depends on ordered neighbor connections. The positive feedback effected by the excitory neighbor connections enhances the features of the input that match patterns embedded in the excitatory synaptic weights. The overall strength of the excitatory response is used to suppress outliers via the dynamical inhibitory threshold imposed by the global inhibitory neuron. Thus the circuit can be seen as imposing an interpretation on an incomplete or noisy input signal, by restoring it towards some fundamental activity distribution embedded in its excitatory connections \cite{Hahnloser00,Hahnloser03}. This selective amplification can be steered in an attentional-like manner by introducing 'pointer neurons' in the excitatory feedback loop, which bias the WTA so that activity at the preferred location 'wins' even if the activity at another competing location is larger \cite{Hahnloser99}.

One drawback of the sWTA is that it is (for the parameters used here) not hysteretic. The activity of the network relaxes toward zero when the input is removed. However, to perform useful computation, the reaction of a network to the same input should depend on the pattern of previous inputs. Such state-dependent processing requires a hysteretic element that is able to retain a history of previous states. One way to provide this history is by inserting into the network attractors that remain stable in the absence of input. We show how pairs of sWTA circuits can be configured to provide multiple states, which are different patterns of sustained discharge. These states are stable in the absence of external input, and transitions between states are driven by external signals. This property enables the network to react differently to the same external signal depending on the current state. We demonstrate this property by implementing a neural version of a Discrete Finite Automaton.

A Discrete Finite Automaton (DFA) is a computational device that implements state dependent processing of strings of input symbols. It comprises a set of states (nodes); a transition function that describes the transitions (edges) between states and their dependence on specific input symbols; and a list of acceptable input symbols. The transition between the current state and one of the allowable next states depends on both the current input symbol and the current state. That is, the processing of input symbols is state dependent. This DFA model can be used to define a 'language' by deciding which input strings of symbols lead to a particular final state (the accept state). The set of all acceptable strings is defined as the language of that DFA. DFAs can implement any language belonging to the class of 'regular' languages \cite{Hopcroft00}. 

In the following, we describe simple rules to systematically construct an arbitrary DFA using nearly identical recurrent maps. The significance of this work is that it offers a method whereby the cortex could achieve a broad range of sophisticated processing by only limited specialization of the same generic neuronal circuit.

\section{Results}

\begin{figure}
\centering
\includegraphics[angle=0,width=\linewidth]{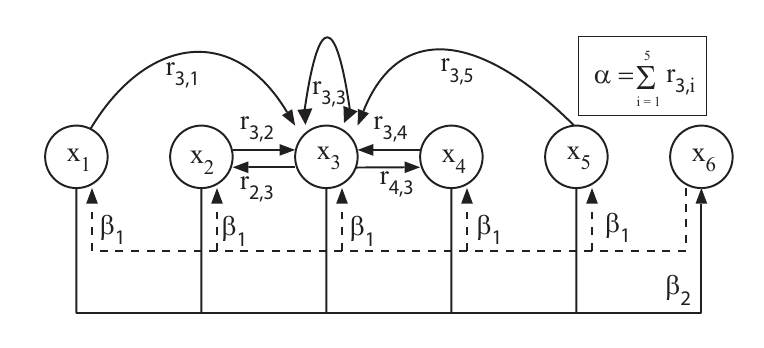}
\caption{
Structure of a single recurrent map that is composed of a number of excitatory units (here $x_1$-$x_5$), and one inhibitory unit (here $x_6$). Each excitatory unit projects to and receives input from the inhibitory neuron. Each excitatory unit is also connected symmetrically to its neighbors as well as itself (connections are shown for the example of $x_3$ and $x_6$). Note that the inhibitory unit $x_6$ does not connect to itself (no self-inhibition).}
\label{figSingleMap}
\end{figure}

\subsection{Single recurrent map}
The behavior of the neural state machine depends on the properties of 
the recurrent map, and so we begin by reviewing these (see also \cite{Abbott94,BenYishai95,Hahnloser00,AbbottDayan01,Douglas2007_recurrent}). Our recurrent map $\mathbf{x}$ consists of $N$ neurons with continuous-valued outputs. $N-1$ of these neurons are excitatory ($x_{1..N-1}$) and one, $x_N$, is inhibitory (Fig \ref{figSingleMap}). Each excitatory neuron receives excitatory input from itself, its neighbors, and a common inhibitory input ($\beta_1$). Each excitatory neuron projects to the inhibitory neuron with strength $\beta_2$. The inhibitory neuron does not connect to itself.

For convenience, we choose a firing rate model \cite{AbbottDayan01} for the recurrent map neurons. Then, in the simple case of self-excitation of strength $\alpha$ (see Fig \ref{figSingleMap}), the dynamics of each excitatory neuron $i$ on the map is given by 
\begin{equation}
\tau \dot{x_i} + x_i = f( I_i + \alpha x_i - \beta_1 x_N - T_i)
\label{eq:recmapE}
\end{equation}
and the dynamics of the inhibitory neuron $j$ is given by
\begin{equation}
\tau \dot{x_N} + x_N  = f( \beta_2\sum_{j=1}^{N-1} x_j - T_N)
\label{eq:recmapI}
\end{equation}

$I_i$ is a constant external input to unit $i$, which is usually $I_i=0$; and $\tau=1$. The firing rate activation function $f(x)$ is a non-saturating rectification non-linearity $max(0,x)$. We also tested a non-linearity of the form $log(a + \exp(b (x+c)))$, where ($a$,$b$ and $c$ are constants), and obtained very similar results. This non-linearity has the benefit that it is continuously differentiable, which is necessary for the analytic analysis of fixed-points. We will interpret the thresholds as possible control inputs, and so they are expressed as arguments $f$. However, these thresholds are usually the same for all units $T=T_i=T_N$. 
All integration was performed with Euler integration with $\delta=0.05$, unless specified otherwise.

\subsection{Amplification by single recurrent map}
\begin{figure}
\centerline{\includegraphics[angle=0]{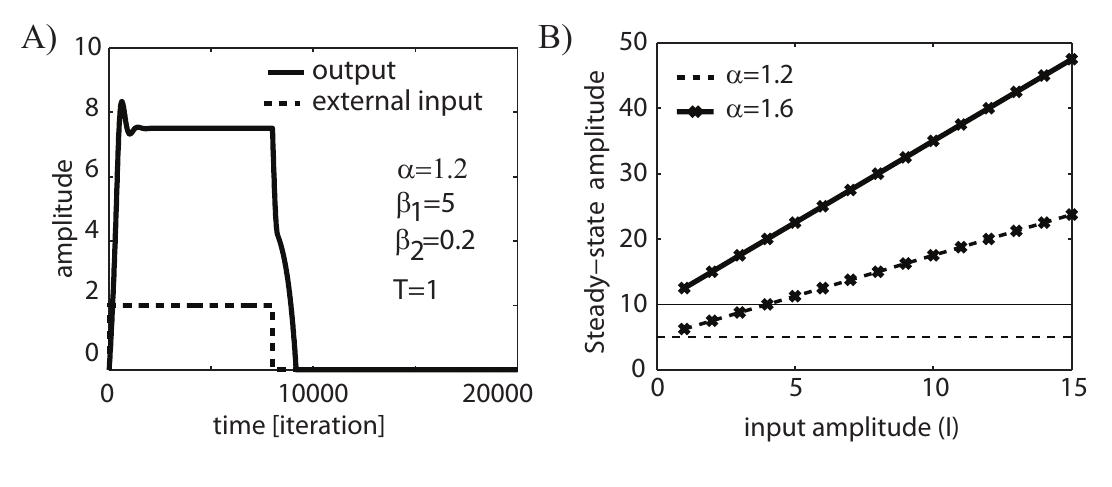}}
\caption{{(A) Following application of an external input to one excitatory unit, the network activity increases to steady amplitude. This activity relaxes back to zero after the input is removed ($\alpha=1.2$, $\beta_1=5$, $\beta_2=0.2$ and $T=1$). (B) The steady state amplitude of the output (bold lines) as a function of the input. The amplitude of the output consists of a constant offset (provided by recurrence, flat lines) and amplification (slope of curve) of the input. Note that the input amplitude needs to be $I>T$ for the network to be activated.
}
}
\label{figSimpleRecc}
\end{figure}

Over a broad range of parameters, the recurrent map will linearly 
amplify a constant external signal (see appendix). The network is continually sensitive to its input, in the sense that its output will relax toward zero when its input is removed (Fig \ref{figSimpleRecc}A). The output follows the input over time, provided that $I(t)>0$. The amplitude of the steady state output is the sum of a variable and a constant component (Fig \ref{figSimpleRecc}B and Eq\ref{eq:steady2Simplified}). The variable component is the input $I$ amplified by gain $G = \frac{1}{1 + \beta_1 \beta_2 - \alpha}$ (the slope of Fig \ref{figSimpleRecc}B). The constant component of the response (flat lines in Fig \ref{figSimpleRecc}B) is independent of the input current (provided that the input threshold is exceeded). However, it does depend on $G$, as well as the threshold and the inhibitory coupling, $\beta_1$.

\begin{equation}
x_i = \frac{I_i}{1 + \beta_1 \beta_2 - \alpha} + \frac{T(\beta_1-1)}{1 + \beta_1 \beta_2 - \alpha}
\label{eq:steady2Simplified}
\end{equation}.

\subsection{Combining two recurrent maps}
\label{section:combRecMaps}

\begin{figure}
\centerline{\includegraphics[angle=0]{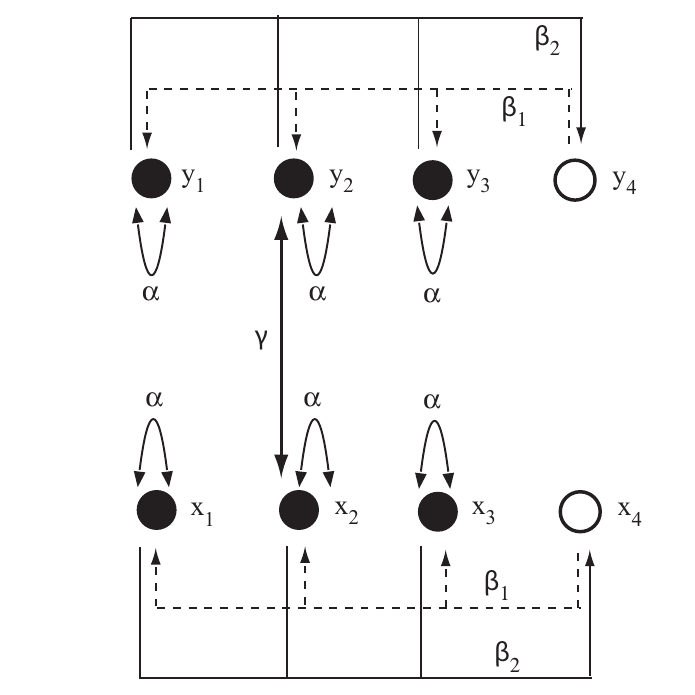}}
\caption{{Two recurrently coupled maps. Excitatory neurons are black, inhibitory neurons white.}}
\label{fig:recurrentCoupled}
\end{figure}

A single recurrent map consists of multiple inhibitory and excitatory feedback loops and it can thus, in principle, give rise to oscillatory or chaotic activity \cite{strogatz94,Wolfram84}. 
Here, we focus on a range of parameters that result in a steady state whenever the external input is constant (for reasons that will become clear later). 
The activity of such a single recurrent map relaxes to zero after the input is removed. However, when two recurrent maps are combined by some simple recurrent coupling, then their activity can be sustained even in the absence of input (see also \cite{Xie2002_double-ringa,Zhang96}). For example, we combine two recurrent maps $x$ and $y$, each of length $N$ (Fig \ref{fig:recurrentCoupled}) and local connectivity described by the $N$x$N$ weight matrices $R^x$ and $R^y$, respectively. Each excitatory unit is connected to its neighbors with $R_{ij}=\exp{(-\sigma d^2)}$ where $d$ is the distance between postsynaptic unit $i$ and presynaptic unit $j$. Here, $\sigma=1$. Each excitatory unit also drives the single inhibitory neuron of its map with $R_{Ni}=\beta_2$, and receives input from that inhibitory neuron with $R_{iN}=-\beta_1$. 

The performance of the recurrent maps depends on the loop gains for excitation and inhibition (see Appendix). So, we simplify the calculation of the excitatory gain by normalizing to $\alpha$ the excitatory connections received by any unit. That is, $\sum_{i=1}^{N} R_{ij}=\alpha$, so that $\alpha$ now sets the total potential excitatory strength received by post-synaptic neuron $i$ from all presynaptic neurons $j$.

The connectivity between the two maps is described by a symmetric weight matrix $C$. 
The weights of excitatory connections between units $k$ on both maps are set to $C_{jj} = \gamma \exp(-\sigma d^2)$ for $j=1..N-1$, where $d$ is the distance between unit $k$ and $j$. The inhibitory neurons of the two maps are not connected ($C_{NN}=0$). 

The $R$ and $C$ matrices of the pair of maps are combined to form the overall weight matrix $W$

\begin{equation}
W = \left[ \begin{array}{cc}
R^x & C \\
C & R^y
\end{array} \right]
\label{eq:W}
\end{equation}

Thus, $W$ describes both the local connections of each map, as well as 
their interactions (for example, Fig \ref{figStateSwitch}B). Using $W$, 
the dynamics of the entire system can be described by 

\begin{equation}
\tau \dot{\mathbf{z}} + \mathbf{z}  = f( \mathbf{u} + \mathbf{p} + W\mathbf{z} )
\label{eq:recmap}
\end{equation}

where bold and uppercase letters indicate vectors and matrices respectively. The vector $\mathbf{z}=[\mathbf{x} \mathbf{y}]$ describes the activity of all units of both maps, $\textbf{u}$ is the external input. The input $\mathbf{p}$ is from transition neurons that will be described later.

\subsection{Stable memory state with two coupled maps}

When two recurrent maps are recurrently inter-connected by excitatory neurons with symmetric weights $\gamma$, there is a range of conditions (see appendix) 
that permit the overall network to retain stable, non-zero, states in the absence of input. The amplitude of the memory state of the excitatory units is given by

\begin{equation}
x_i = \frac{T(\beta_1-1)}{1+\beta_1 \beta_2 - \alpha - \gamma}
\label{eq:steady5Main}
\end{equation}.

This expression is similar (except for $\gamma$) to the constant offset term described above for the single map (Eq \ref{eq:steady2Simplified}). A memory state exists if $T>0$ and $\beta_1 > 1$. From the steady state alone it may appear that it is sufficient to have only one map (equivalent to $\gamma=0$). However, $\gamma>0$ is required for reasons of the dynamics: only a non-zero value assures that the memory state is also an attractor (see Appendix).

The amplitude of the memory state depends on the product of the gain of the network (compare Eqs \ref{eq:steady5Main} and \ref{eq:steady2Simplified}) and the threshold $T$. This perplexing result can be clarified by decomposing $T$ into two different thresholds $T_{exc}$ and $T_{inh}$ for the excitatory and inhibitory units respectively. Then the steady state potential becomes

\begin{equation}
x_i = \frac{\beta_1 T_{inh}-T_{exc}}{1+\beta_1 \beta_2 - \alpha - \gamma}
\label{eq:steady6Main}
\end{equation}.

Now it becomes clear that the memory state depends critically on an inhibitory threshold: provided $\beta_1>\frac{T_{exc}}{T_{inh}}$, the memory state exists only if $T_{inh}>0$, whereas the excitatory units can have a zero threshold $T_{exc}=0$ (Eq \ref{eq:steady6Main}). $T_{inh}$ is effectively thresholded disinhibition. This means that recurrent excitation can grow with high gain, until the inhibitory threshold is exceeded, so that the response of the network is stabilized. 
This explains why the inhibitory threshold controls the amplitude of the memory state. 

\begin{figure}
\centerline{\includegraphics[angle=0,width=\linewidth]{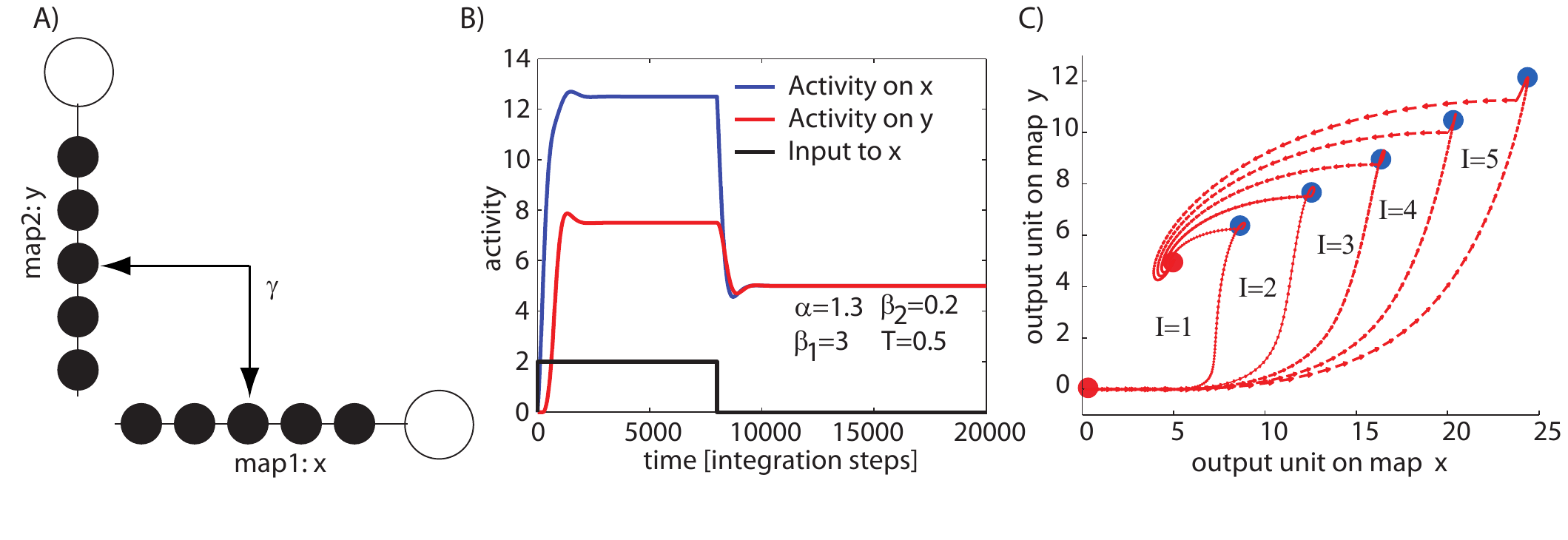}}
\caption{{Two connected recurrent maps maintain stable states in the 
absence of external input. \textbf{A)} Illustration of connectivity. 
Maps are recurrently connected by $\gamma$. Each map has one inhibitory 
neuron (big circles). One neuron on each map (here $x_3, y_3$) 
are symmetrically connected to one another with weight $\gamma=0.1$. 
\textbf{B)} Input is applied to a unit on map $x$ that is recurrently connected to map $y$ (here $x_3$).  After offset, the network relaxes to the memory state. \textbf{C)} Dynamics of the two units $x_3$ and $y_3$, shown in phase space for different input amplitudes, applied to neuron $x_3$. The red dots denote points ($\dot{\mathbf{x}}=0$ and $\dot{\mathbf{y}}=0$) that are stable attractors in the absence of input. The blue dots are stable attractors in the presence of input. After the input of any amplitude is withdrawn, the network converges to the common red attractor.  The vectors indicate the value of the derivative $\dot{x_3}$ and $\dot{y_3}$. Here, $\delta=0.01$ (numerical integration). 
}}
\label{figTwoMaps}
\end{figure}

The memory state property is demonstrated in Fig \ref{figTwoMaps}B, using a network with the following parameters: $\alpha=1.3$, $\beta_1=3$, $\beta_2=0.2$, $T=0.5$ and $\gamma=0.1$. In this example, for simplicity, only self-excitatory connections are used within each map. And, we have plotted only the two units on maps $x$ and $y$ that are recurrently connected ($x_3$ and $y_3$). 

While the external input is applied, the $x$ unit output is equal to the anticipated steady state activation (\ref{figTwoMaps}B) described by Eq \ref{eq:steady2Simplified}. The output of the $y$ unit is smaller because it does not receive an external input: Its response arises from the input it receives from other units in the overall network.  When the driving input is removed, the network relaxes to a non-zero stable state where $x_i=y_i$ and $x_N=y_N$ (excitatory and inhibitory neurons, respectively). Thus, the network exhibits state (or memory) by maintaining persistent (and constant) output. The amplitude of this state is described by Eq \ref{eq:steady5Main}. The amplitude of the sustained response is independent of the amplitude of the input: The same memory state is reached after various amplitudes of input are removed (Fig \ref{figTwoMaps}C).

\subsection{Robustness to noise}
\label{sec:robustness}

\begin{figure}
\centerline{\includegraphics[angle=0,width=\linewidth]{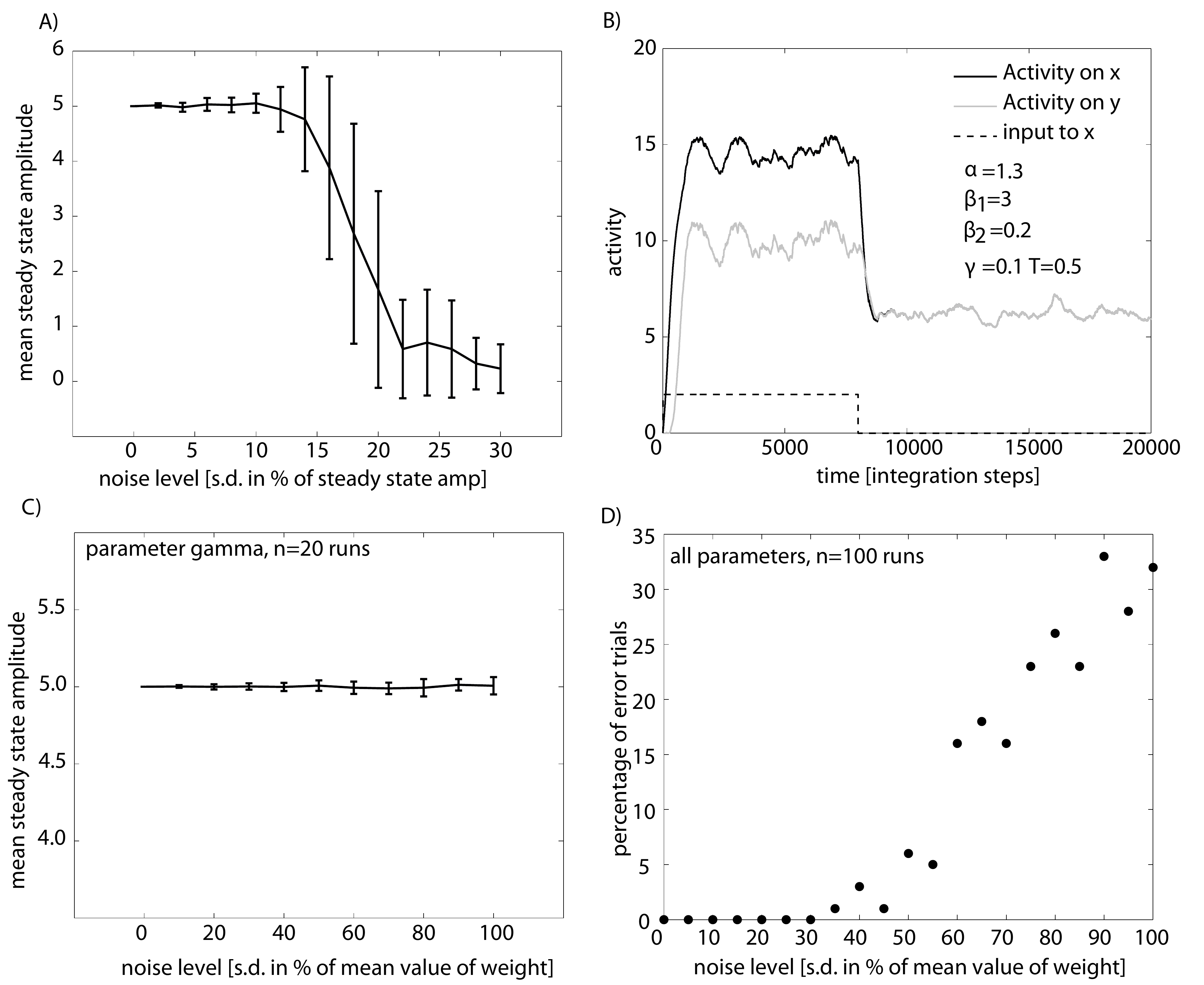}}
\end{figure}
\begin{figure}
\caption{{Stable states can be maintained in the presence of noise. All parameters are equal to Fig \ref{figTwoMaps}.
\textbf{A)} Mean steady state amplitude as a function of the amplitude of readout noise.
The noise amplitude (s.d.) is specified relative to the steady state amplitude. The network can tolerate up to 15\% readout noise.
\textbf{B)} Network activity in the presence of weight noise ($\gamma$, s.d. of noise equal to 0.1). Compare to Fig \ref{figTwoMaps}B. 
\textbf{C)} Mean steady state amplitude as a function of weight noise. In this example, noise was only added to $\gamma$. 
The network can tolerate noise amplitudes (s.d.) of up to 100\% of the original weight of $\gamma$. Here, $\delta=0.01$ (numerical integration).
\textbf{D)} Percentage of trials where the memory state was maintained successfully as a function of the amplitude of noise applied to all weights of the network
simultaneously. The network can tolerate noise with a standard deviation of up to 30\% of the nominal value of the weights. 
}}
\label{figNoise}
\end{figure}

Recurrently connected networks can be very sensitive to noise, particularly when these networks lack inhibition. However, recurrent maps (which by definition have inhibition) are very robust against noise. We confirmed this robustness in the coupled maps by introducing two kinds of signal variability: Readout (output) and synaptic (weight) noise. Output noise was added by an additional term in Eqs \ref{eq:recmapE} and \ref{eq:recmapI}:

\begin{equation}
\tau \dot{x_i} + x_i = f( \alpha x_i + I_i - \beta_1 x_N - T + \mathcal{N}(0,\sigma))
\label{eq:recmapEnoise}
\end{equation}

We tested the same network as described above (Fig \ref{figTwoMaps}B) while varying the standard deviation of the noise, and determined whether the network reached its memory state by calculating the mean steady state amplitude after the external input was removed (timesteps 10000-20000 in Fig \ref{figTwoMaps}).

First we evaluated robustness to readout noise by adding variable amounts of noise to the output of each unit. The noise for each unit was drawn independently
every $\frac{\tau}{10}$ timesteps. The exact sampling interval of the noise is not critical, provided that it is significantly larger than the integration time-constant and significantly smaller than $\tau$. The results show that the network can tolerate readout noise with a standard deviation of up to approximately 15\% of the steady state amplitude (Fig \ref{figNoise}A). If the noise amplitude exceeds this value, the network relaxes to zero instead of the memory state after removal of the input. Note that readout noise is also equivalent (Eq \ref{eq:recmapEnoise}) to random variability of the threshold $T$.

Next, we evaluated robustness to synaptic noise. Weight noise was added to the weights by setting $j = j + \mathcal{N}(0,\sigma)$ for $j \in [\alpha,\beta_1,\beta_2,\gamma]$. New values of the noise were drawn every $\frac{\tau}{10}$ timesteps from a truncated normal distribution (bounds were set to $\pm j$ for $j \in [\alpha,\beta_1,\beta_2,\gamma]$ to prevent negative weights). We varied the standard deviation of the noise ($\sigma$) and quantified robustness.
$\sigma$ was set such that it equals a certain percentage of the nominal value of the parameter value. For example, if noise is set to 10\% and the nominal value of
the parameter under investigation is $2.0$, the standard deviation of the noise equals $0.2$.

First, we evaluated synaptic noise by adding noise to one of the 4 weights $\alpha, \beta_1, \beta_2, \gamma$ only. Note that this description refers to the
type of weight and not an individual weight. Thus, for example, if noise was only added to $\alpha$, different noise values were added to each instance of $\alpha$
in the entire network.  For demonstration, we focus here on the weight $\gamma$ but similar results apply to the other weights. We found that the network can tolerate weight noise with a standard deviation of up to 100\% of the original (noiseless) $\gamma$ value (Fig \ref{figNoise}B,C). Similar amounts of noise were tolerated by the remaining weights.

Next, we added synaptic noise to all weights simultaneously. Independent noise was added to each instance of each weight. We found that the network is very robust to
noise. Running 100 trials for each level, the memory state was stable for all trials up to 30\% (Fig \ref{figNoise}D) noise. 
However, even for higher noise levels (i.e. 60\%) less than 10\% of trials failed (Fig \ref{figNoise}D). Thus the network can tollerate up to 30\% noise on all weights.

An other source of synaptic noise not explicitly tested here is frozen weight noise. That is, the weights differ from their specified values but this difference
does not change over time. This kind of noise is particularly relevant for implementation of circuits in analog hardware \cite{Pavasovic94,SerranoGotarredona99}. 
Typically, the effective weight is approximately 30-50\% of the nominal (specified) weight \cite{Neftci08}. This noise is introduced at time of fabrication and is fixed. 
Our circuits are robust to such noise as long as the resulting parameters are still within the valid range for stability (see appendix). Robustness to such noise can be
increased by choosing nominal parameter values which lie in the  middle of the permitted ranges rather than on the edge.

\subsection{Controlled transitions between multiple memory states}

A DFA consists of nodes and edges (representing the states and transitions, respectively). So far we have only described how coupled maps can be used to construct a network with multiple stable attractors that can implement the DFA nodes. Now we will show how the transitions between these embedded states (the DFA edges) can be induced by transiently applying a short pulse of input to a currently unexpressed target state. That input causes the network to switch from the current state to the target (pulsed) state. 

\begin{figure}
\centerline{\includegraphics[angle=0]{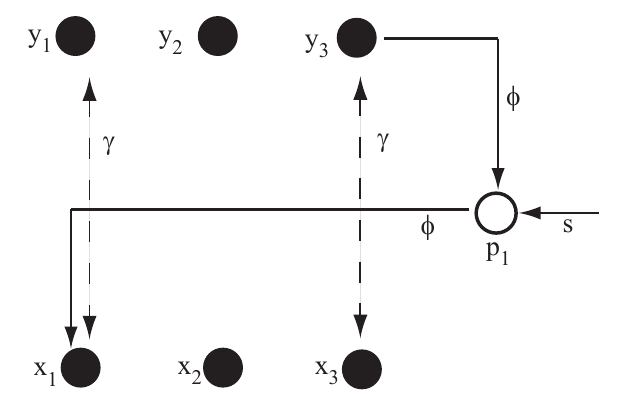}}
\caption{{Transition neuron ($p_1$) that, when activated by input $s$, leads to a switch to a different memory state. 
Only connections between the two maps are shown. 
All local connectivity ($\alpha,\beta_1,\beta_2$) is as shown in Fig \ref{fig:recurrentCoupled}.}}
\label{fig:recurrentTNs}
\end{figure}

We implemented this transition mechanism using a variant of the 'pointer' neurons that have been used previously to bias the competition on a single recurrent map \cite{Hahnloser99}, for example to enable the winning stimulus to emerge at a preferred direction rather than the point of maximal input. A pointer neuron is a unit that is symmetrically connected to the excitatory units on the recurrent map. Thus, the pointer neuron both receives input from and sends output to the excitatory units. This connectivity of the pointer neuron is non-uniform. Previously, a gaussian connectivity profile was used to selectively bias activity in a particular region of the excitatory map \cite{Hahnloser99}. Here, we apply the same basic idea to implement more specialized 'transition neurons' (TNs, Fig \ref{fig:recurrentTNs}, Eq\ref{eq:pointers}). These units temporarily bias the competition such that the currently maximally active unit (the current state) loses the competition in favor of the next (target) state. They combine activity from map $y$ with the external input $s$.  If the appropriate members of $y$ and $s$ become active, the transition neuron becomes active and initiates the required state transition.

\begin{equation}
\tau \dot{\mathbf{p}} + \mathbf{p} = f( P\mathbf{y} + \mathbf{s} - T_p)
\label{eq:pointers}
\end{equation}

$P$ describes the connectivity between the $M$ TNs and the activity of the $y$ map. It has dimensionality $M$x$N$, where $M$ is the number of transitions (edges) of the DFA and $N$ the number of excitatory units on the map. TNs receive no recurrent input ($P_{ii}=0$). Each TN $i$ receives input from the excitatory unit $j$ that represents the state from which the transition originates ($P_{ij}=\phi$). Similarly, the TN sends its output to the unit $k$ that represents the state where the transition leads to ($P_{ik}=\phi$).
The range for possible values of $\phi$ is derived in the appendix.

$T_p>0$ is a constant threshold that suppresses TN output in the absence of external input. The external input $s$ represents the input symbols. Every TN that represents the same symbol receives the same input. A short pulse of activity on that input signals the arrival of that particular symbol. $T_p$ is set equal to the steady state value reached by the map units when external input is applied (see below).

\begin{figure}
\centerline{\includegraphics[angle=0,width=\linewidth]{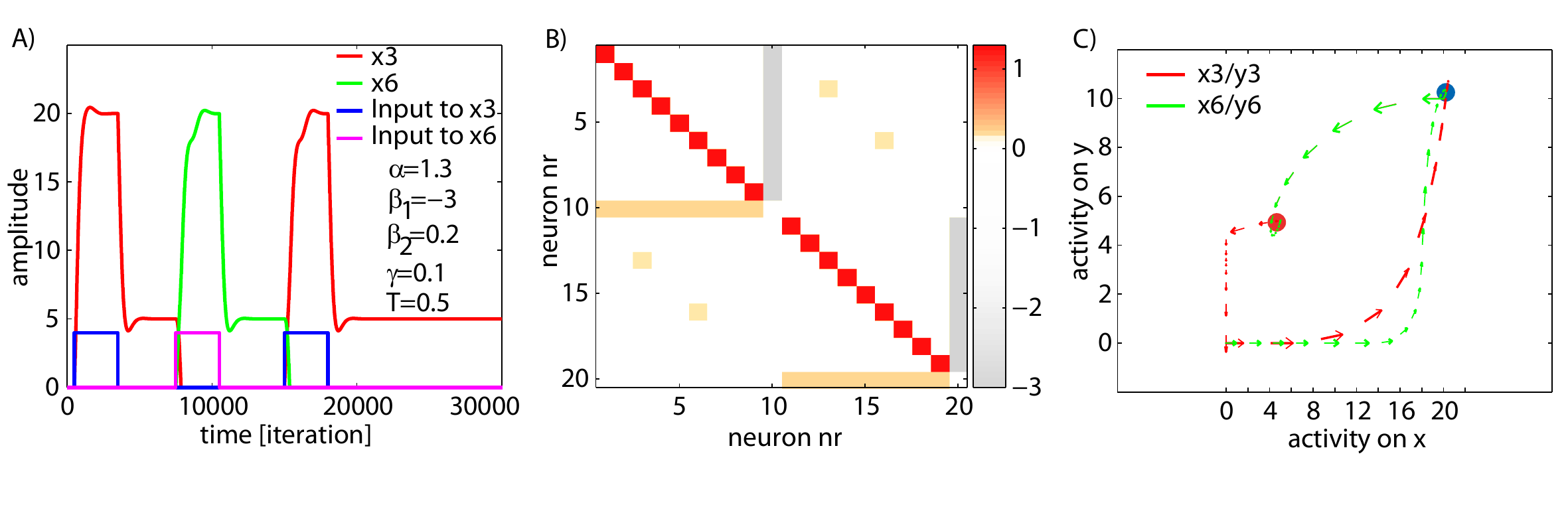}}
\caption{{Short bursts of input induce rapid transitions to different 
states. Two units ($x_3$ and $x_6$) are recurrently connected . 
\textbf{A)} A short burst of input to either $x_3$ (blue) or 
$x_6$ (magenta) elicits a rapid transition. The peak of activity 
remains at the same location after removal of the input (the memory 
state). \textbf{B)} The weight matrix $W$ of 
the network. Positive weights are shown in color, negative weights in 
gray scale. The recurrent excitatory connectivity can be seen on the main 
diagonal. The connections between the map can be seen on the upper and 
lower lines parallel to the diagonal. 
\textbf{C)} Phase space representation of the dynamics of the 
first switch (red to green). The blue and red dots indicate steady 
state while input is applied and in the absence of input, respectively.}}
\label{figStateSwitch}
\end{figure}

We demonstrated this transition mechanism in a network composed of two recurrent maps of $N=10$ units, in which we embedded two states by recurrently connecting units $3$ and $6$ of each map with weights $\gamma$. Thus $C_{jj}=\gamma$ for $j\in[3, 6]$. The other parameters are indicated in Fig \ref{figStateSwitch}A. The network (Fig \ref{figStateSwitch}B) has two stable attractors with peaks of activity at $x_3$ and $x_6$. 

The behavior of the network is shown in Fig \ref{figStateSwitch}A. At first the network is quiescent. After a short initializing pulse of activity to $x_3$, the network relaxes to a peak of activity at unit $x_3$ that is sustained even after the input has been withdrawn. A short pulse of input applied to $x_6$ elicits a transition from the current state $x_3$ to the target state, $x_6$. The state switch occurs because of the competitive nature of the WTA. Transiently, the total input to $x_6$ is larger than to $x_3$. Because of this, the WTA selectively amplifies $x_6$ and suppresses $x_3$. The amplitudes of the steady states during application of the input as well as during the memory period are described by Eqs \ref{eq:steady2Simplified} and \ref{eq:steady5Main}.

Thus, two recurrent maps can be used to construct stable attractors that provide persistent output in the absence of input. For convenience we have shown here only the simplest case of two states. However, additional states can be embedded easily by inserting the necessary additional weights in the connection matrix $C$ (lower left and upper right quadrants in Fig \ref{figStateSwitch}B).

\subsection{Implementation of a state automaton using coupled recurrent maps}
The multiple stable states embedded in the coupled maps, and the transitions between them, can be utilized to implement a DFA. The DFA operates as follows: Starting in an initial state, it reads a sequence of input symbols. For each successive symbol in the input sequence, the DFA transitions to a next state that is determined by the DFA's transition function. The state that the DFA finds itself in when the input is exhausted determines the DFA's evaluation of the input sequence. If on exhaustion of input, the DFA is in an 'accept' state, then the input sequence is deemed 'accepted'; otherwise it is deemed rejected \cite{Hopcroft00}.

We implemented the neuronal DFA as follows. First, we embed the same number of attractors as the DFA has states. To achieve this, we created
two recurrent maps consisting of $m N+1$ units each ($m$ is the number of states, $N$ the number of units per state).
The number of units $N$ required to represent a state is determined by the width $\sigma$ of the
connection profile between the two maps (Eq \ref{eq:W}). For each state 
in the DFA, one pair of units is designated to represent that state 
(here, for a simple example of only two states, we added $N=3$ units per state; the state $q0$ is represented by $x_3, y_3$ and $q1$ by $x_6, y_6$).  Each unit that represents a state is recurrently connected to its counterpart on the other map (here $x_3$ to $y_3$ and $x_6$ to $y_6$) with weight $\gamma$. The lateral connectivity on each map is the same as detailed above.

Next, a transition neuron (TN) is added for each possible state transition. Each TN receives two inputs: one external (the input symbol that initiates the transformation) and one internal (the state from which the transition originates). Either or both of these inputs may be zero, in which cases the output of the TN is bound to be zero so that no transition will occur. To achieve this behavior, each TN receives a constant negative (inhibitory) input $-T_p$ that prevents it from becoming active in the absence of input, even when the network is currently in the preferred state of the current TN. This arrangement also prevents a state transformation if a symbol is received for which there is no defined transition from the current state.

The output of the TN is connected to the unit on map $x$ that represents the target state that the transition leads to. While activated, the TN output will bias the competition on the coupled maps, encourgaing the new state to become dominant. All TNs that represent the same symbol receive the same input. 
Here, the amplitude of the external input to the TNs is set equal to $T_p$. The value of $T_p$ is chosen such that is somewhat larger than the steady-state amplitude reached by the 
map neurons in the presence of constant input (any value larger than $x_i$ as calculated by Eq \ref{eq:steady2Simplified} is allowed). 
Thus, the overall network has as many inputs as there are distinct symbols. The size of the network (in terms of number of units) scales as $O(m+n)$, where $m$ is the number of states and $n$ the number of state transitions. Thus, the network scales linearly with the number of states. While it might be possible to represent states by a combination of
units (and thus scale better), we opted here for an explicit representation of each state for purposes of analysis. 

\subsubsection{Example DFA}
\begin{figure}
\centerline{\includegraphics[angle=0,width=\linewidth]{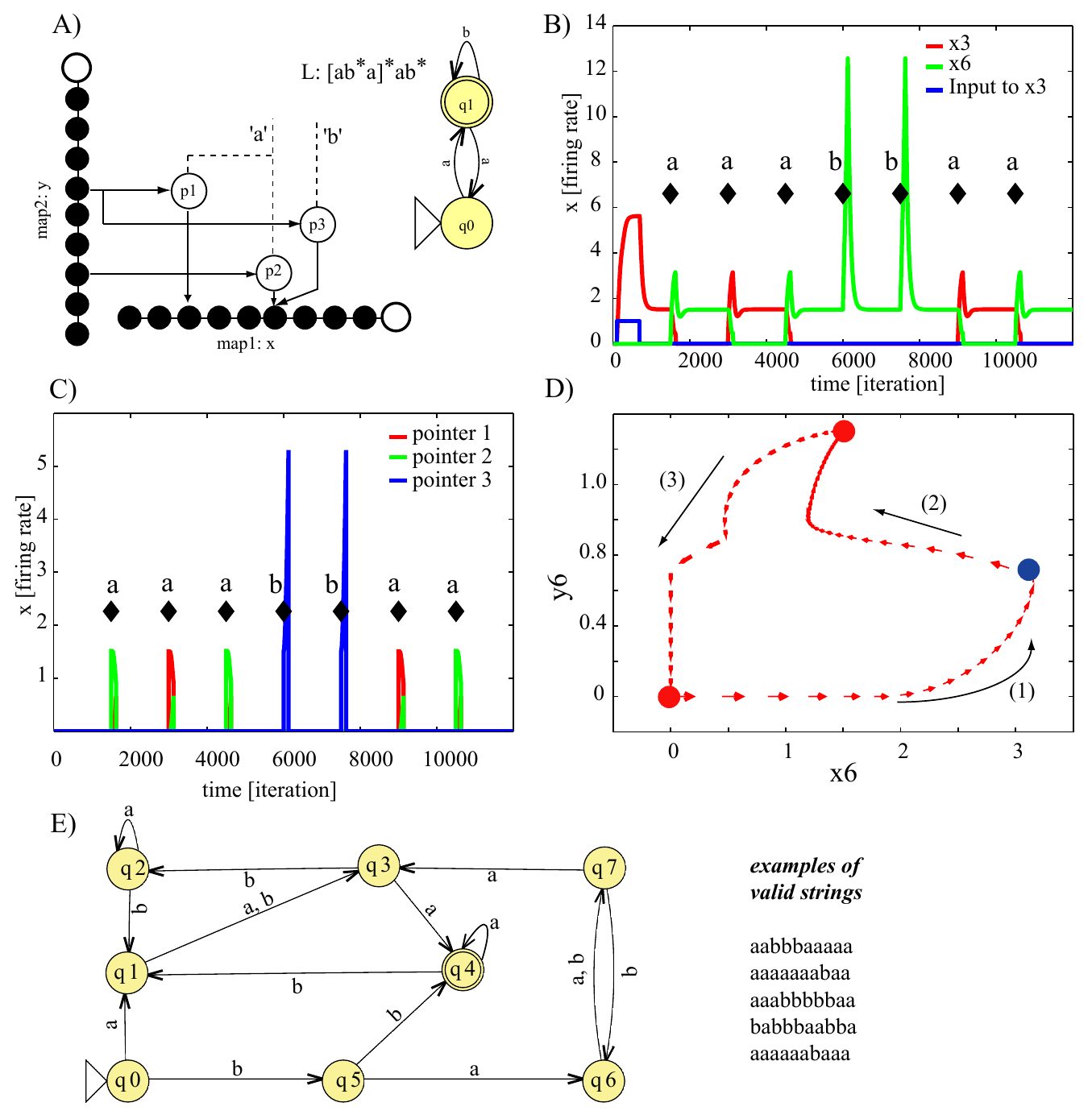}}
\end{figure}
\begin{figure}
\caption{Implementation of a DFA using recurrent maps. \textbf{A)} DFA that implements the language (ab*a)*ab* (right) and its recurrent map implementation (left). \textbf{B) and C)} Processing of the sequence $aaabbaa$. After a short burst of initialization (blue), no external input is applied to the network except for the symbols. The recurrent net switches to the appropriate state (B) and the appropriate pointer neuron is activated (C). \textbf{D)} Illustration of the state switch from $q0$ to $q1$. Activation of the pointer neuron induces a transition to the next state (1). After offset of the pointer neuron input, the network settles to the stable state $q1$ (2). After the network transitions to another state, activity returns to the baseline state (3). \textbf{E)} Example of a randomly generated minimized DFA with 8 states. Example strings that lead to the accept state are shown on the right.}
\label{fig4}
\end{figure}

We demonstrated these properties by constructing a simple DFA that implements the regular language {\it[ab*a]*ab*} (Fig \ref{fig4}A) The coupled map has two states ($q0, q1$). Its input alphabet are the two symbols $a$ and $b$ and it has the transition functions $q_1=(q_0,a)$, $q_0=(q_1,a)$ and $q_1=(q_1,b)$. The example input string {\it aaabbaa} is a valid string under this language and brings the DFA to the accept state $q1$ (represented by $x_6$). In this more sophisticated example of a coupled map, neighbor connections are used between the local excitatory neurons, and also between the cross-coupled populations.

The dynamics of $x_3$ and $x_6$ during the processing of the input 
string confirms that the neuronal DFA executes each state transition correctly (Fig \ref{fig4}B). The dynamics are further illustrated by plotting the activity on map $y$ as a function of the activity on map $x$ (phase space, Fig \ref{fig4}D). Each unit $x_i$ that represents a state of the DFA has two stable attractors (where $\dot{x_i}=0$), in the absence of input (Fig \ref{fig4}D) (red dots). Before processing the input string, the DFA must be initialized to its initial state by a short external pulse (blue in Fig \ref{fig4}B). Note that such neurons, which indicate the start of a sequence, have actually been observed in the cortex
(see discussion). Thereafter, short pulses of input are applied to the appropriate TNs, depending on whether the symbol $a$ or $b$ is present (black diamonds in Fig \ref{fig4}B,C). 
The duration of the external inputs, representing the 'symbolic' inputs, needs to be long enough for the network to converge. Here, we used $15\tau$ (300 iterations for a $\delta=0.05$.
The number of iterations necessary for the network to execute a state switch as well as to relaxe after removal of the external input does not depend on the number of states (see appendix for a calculation) nor on the number of TNs. However, it does depend on the parameters of the network. 

\subsubsection{Implementation of DFAs of arbitrary size}
So far we have demonstrated how to implement a DFA with 2 states and 4 transformations using the coupled maps. Is it possible to implement DFAs of arbitrary size using the same construction rules? To facilitate the construction of arbitrary DFAs, we have developed software that automatically converts a DFA constructed in a graphical utility for constructing DFAs (JFLAP) \cite{Rodger06} into two dynamically coupled maps \footnote{Source code is available on the first authors webpage or can be requested by e-mailing the authors.}. We used this software to generate random DFAs using the method described in \cite{Bongard05}. 
We used a standard algorithm \cite{Hopcroft71,Hopcroft00} to minimize DFAs with respect to the number of states used to represent the language represented by the random DFA.
The random DFAs were automatically converted to a weight matrix for the two WTAs and their interactions (the states), as well as the connections of the TNs.

The weight matrix was constructed using the following rules: i) add a few units (here, we used 5) to both maps for each state; ii) add one TN for each transformation and connect them appropriately and iii) connect each state in the weight matrix as described above. 
How many units are added per state (here 5) depends on the width of the 
local connectivity (see $\sigma$ in section \ref{section:combRecMaps}; the bigger $\sigma$, 
the more units per state need to be added such that the activity bumps 
of two states do not overlap). 
The generated DFA was then evaluated in terms of its ability to correctly classify randomly generated strings. Like the standard DFA, the coupled map DFA was deemed to classify a string correctly (or, 'recognize' the string) if, after all input symbols have been processed, the DFA reaches its accept state. If any other state was reached, the processing was incorrect (unless the input was not a string of the language). We tested random DFAs of up to 40 states with 100 random strings each, and found that all these strings were processed correctly (an example is shown in Fig \ref{fig4}E).

\subsubsection{Robustness}
Long chains of excitatory neurons are capable of producing sophisticated patterns of activity, as exemplified by synfire chains \cite{abeles91}. However, the more neurons follow each other in this feedforward arrangement, the more sensitive the output becomes to the input, because errors accumulate and propagate in those architectures. Does this error accumulation occur in our architecture? If yes, then the network should fail to process long sequences of symbols. While the transition executed for each input symbol only depends on the current state, the current state includes information about all previous input symbols (it fulfilles the Markov
property). Thus, the appropriate test is whether the network can reliably process long sequences of input. If errors accumulate, the network behavior should break down beyond a certain string length. 

\begin{figure}
\centerline{\includegraphics[angle=0,width=\linewidth]{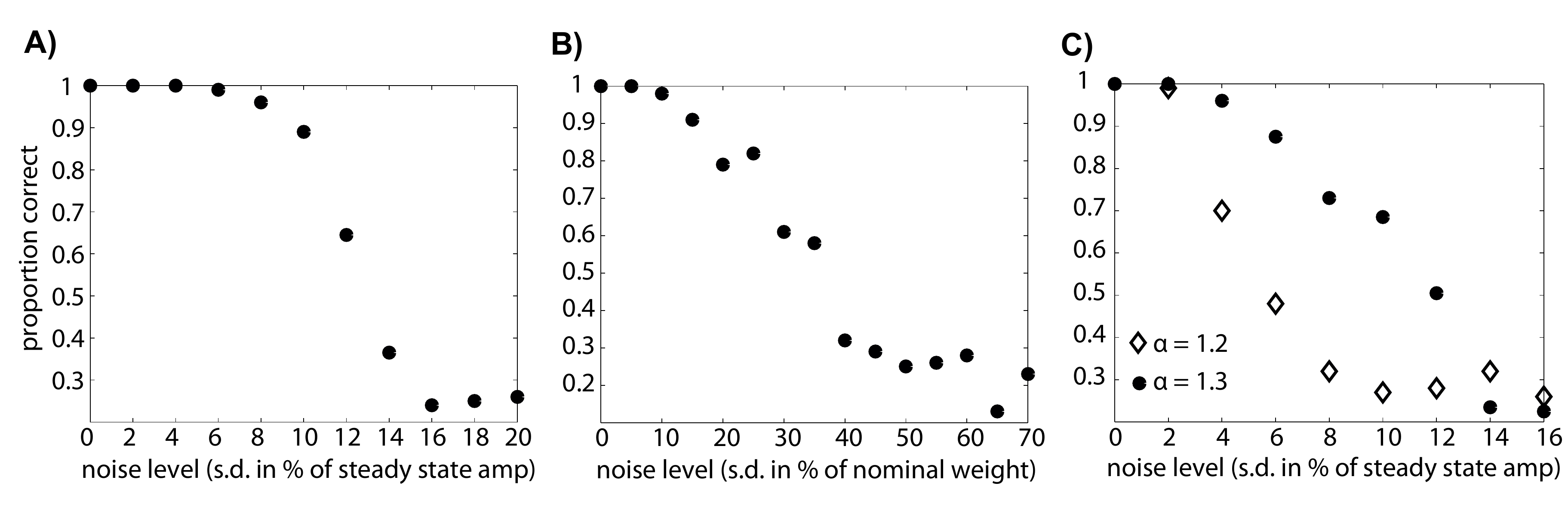}}
\caption{{Illustration of noise robustness of a network implementing a random DFA with 4 states (chance performance is 25\%). The parameters of the network are: $\alpha=1.3, \beta_1=3, \beta_2=0.2, \gamma=0.1,
\phi=0.88, T=0.5$. The integration time constant is $\delta=0.05$. Networks were fed random strings of length 4 to determine whether the network reaches the correct state.
\textbf{A)} Performance for different levels of output noise (applied to all units). The steady state amplitude of the network was 5.0, thus 10\% noise correspond to a s.d. of 0.5. \textbf{B)} Performance for different amounts of weight noise. Noise was applied to all weights. The s.d. of the noise was set relative to the nominal weight of each weight (y-axis). \textbf{C)} Network performance for output noise and different values of $\alpha$ (1.2 vs. 1.3). Note that the larger value of $\alpha$ results in a more noise robust network for this configuration. All other parameters are the same.
}}
\label{figDFAnoise}
\end{figure}

There are two principal actions of the network which are sensitive to noise: keeping the memory state (section \ref{sec:robustness}) and executing a state switch. 
Here, we evaluate both simultaneously. Without the presence of artifical noise sources, 
we found that long random strings (we tested up to 30 symbols) are processed with the same reliability as short strings. Thus, numerical errors of the integration do not
accumulate. Next, we added output or synaptic noise as described in section \ref{sec:robustness}. The performance of one particular network as a function output and 
synaptic noise levels is shown in Fig \ref{figDFAnoise} (see figure legend for network parameters). This particular network can tolerate output noise of 8\% of the steady state amplitude als well as approximately 10\% weight noise. After the noise threshold is reached, performance degrades rapidly and quickly reaches chance performance (4 states, thus 25\% in this case). 
The level of noise that a particular network can tolerate depends on the parameters of the network (i.e., the weights, the number of units per state), the distribution of the noise, and the sampling frequency of the noise (see Fig \ref{figDFAnoise}C for an example). Also, if $\alpha_i>0$ (for $i>1$), robustness to noise depends on the spread of connectivity (determined
by $\sigma$, see section \ref{section:combRecMaps}) as well as the number of units dedicated to a particular state. The bigger $\sigma$, the more units are required to represent
a particular state. Whether this increases or decreases reliability depends on the other parameters of the network. Also, it is beneficial to choose the weights such that they lay
in the middle of the permitted values rather than on the edge, such that noise does not cause instabilities in the network.

This result demonstrates one of the crucial aspects of computations with WTAs: states are stable no matter what the initial conditions. Thus, noise does not accumulate over time because the signal is restored (the noise is rejected) at every step. This is the reason why previous approaches using saturating neural networks (see discussion) can not process long strings: performance degrades even if no noise is added (due to numerical errors). No matter how accurate the feedforward network is, the performance of long string acceptance eventually degrades. In contrast, this problem does not exist for the DFA implemented using WTAs. A similar strategy has also been used to make synfire chains robust to noise \cite{Diesmann99}. 

\section{Discussion}
It has been known since the foundational report of McCulloch and Pitts 
\cite{McCulloch1943_ideas} that recurrently connected neuronal networks  have the properties of a finite state machine. Subsequent work has elaborated and clarified those properties: For example Minsky \cite{Minsky1967_computation} from the point of view of general computation;  Grossberg and Hopfield from stability \cite{Cohen83,Hopfield84}; and Elman and Forcada \cite{Carrasco00,Kremer95,Elman91} from the point of view of network computation. In general, however, these approaches have assumed inherently reliable neurons with saturating 'neuronal' activation functions. By contrast, biological cortical neurons are subject to various sources of noise, and operate well beneath saturation in the linear range of their activation function \cite{Douglas95}.

The contribution of this paper is to demonstrate that robust state-dependent processing can be embedded in networks of neurons whose activation function is both non-saturating and non-linear (thresholded). Consequently, they can depend only on their dynamics of inhibition, excitation and thesholding for their noise resistance and stability. Moreover, we have provided simple rules that permit the systematic design and construction of a (nearly) arbitrary neuronal state machine composed of nearly identical recurrent maps. We claim 'nearly arbitrary' because for practical reasons we tested the automatic design process only out to 40 embedded states (because it is not easy to find a minimized random DFA consisting of more than 40 states); and 'nearly identical' because some specific connections are required to implement a specific state machine within an otherwise generic neural architecture.

Essentially, we construct a multistable network that embeds a number of states, by coupling two recurrent maps whose weights have been set to provide soft winner-take-all (sWTA) performance. The two sWTAs have identical connectivity, except for a small fraction of symmetrical (recurrent) cross-connections between them that are used to embed the required states. This recurrent coupling between the maps allows them to sustain (remember) their current state in the absence of input. 

The network can be switched from its current state to another of its states by applying a short burst of input. The required transitions between states are effected by a class of neurons that activate the new state, conditioned on the current state and also an external input 'symbol'. Thus, the network's reaction to the input is state dependent.

A class of 'transition neurons' drive the state transitions by combining 
the current state of the network (memory) with an input symbol to excite 
the next state of the network. The transition neurons are similar to the 
pointer neurons described by Hahnloser et al \cite{Hahnloser99}, which they used to steer feedback over the entire range of a single map of neurons. For example, pointer neurons are able to bias WTA behavior in favor of inputs at certain locations of the map, so that activity at these locations wins the WTA competition even if they are numerically smaller than activities at others' (which would otherwise win an unbiased WTA competition). In that application, the pointer neurons participate directly in the feedback between the neurons of the WTA, and so participate directly in the ongoing computation of the WTA. By contrast, our transition neurons are only activated when a new input symbol is processed. As soon as the state switch has occurred (or no input symbol is applied), the transition neurons become inactive again. In our network, this selective activation is due to inhibition that is continuously applied to all transition neurons. A transition is triggered only when the combined activity of the current state and the external input exceeds the inhibitory threshold.

The networks described here can function over a wide range of weights 
(as long as certain conditions are met, see appendix for a summary). Of 
particular interest is that the coupling constant $\gamma$ between the 
two recurrent maps must be small compared to the local recurrent 
excitatory 
weights ($\alpha\gg\gamma$). Thus, a very small weight is sufficient to 
couple two independent WTAs such that they bias each other's winner. This is well in line with our previous finding that the sensory excitatory input to pyramidal neurons in layer 4 of visual cortex is minor compared to excitatory input from other neurons of the same layer \cite{Binzegger04}.

Our model uses a single global inhibitory neuron per map to implement 
the competitive component of the sWTA. Which excitatory units compete with each other is determined by whether they share
common inhibition. Experimental evidence exists for inhibitory feedback loops on many different scales, such as
local lateral inhibition and diffuse inhibitory feedback mediated by the thalamus \cite{Douglas2004_neuronal}. Such feedback loops can 
enforce competition and thus result in WTA behavior. In some structures, it has been experimentally demonstrated that such global (diffuse) inhibition exists and that they serve to enforce WTA type competition, i.e. \cite{Kurt08,Baca08,Tomioka05}. WTA networks can, however, also be implemented using more local forms of inhibition (i.e. \cite{Yuille89}).
Global inhibition necessarily makes the weight matrix asymmetric and so it is difficult to derive 
analytically the conditions on the weights that guarantee convergence. One approach to this difficulty has been to assume that locally, inhibition is infinitely fast \cite{Grossberg73,Wersing01}. However, we wanted to keep our network physiologically plausible and so could not accept this assumption. Seung \cite{Seung97} has suggested a Lyapunov function for inhibitory-excitatory networks similar to those used here.  His function requires the inhibition to be symmetric ($\beta_1 = -\beta_2$). However, if a WTA is to function properly, it is crucial that this requirement is not upheld, and so we could not use Seung's approach. Instead, we used linearization at steady states and numerical simulations to confirm that our approach is capable of implementing arbitrary DFAs.

We aim to model state dependent processing of external inputs. Our network thus requires an explicit external input to trigger a transition to the next state. This transition is state dependent. These external inputs could be the result of some action performed at every state and would thus not be available in advance. We used the processing of a list of symbols as an abstract model of this behavior. An other application of sequential transitions between states has been the memorization of sequences. In this case, an external stimulus would trigger the automatic replay of the entire sequence without requiring any further external inputs. An example is winnerless competition, where networks are constructed such that each state is a saddle point with only one unstable direction, which leads
to the next stable point \cite{Selinger03,Rabinovich01}. While this was not our principal aim, our network could also be enhanced to allow autonomous transitions between states. This could be achieved by replacing the TNs with units that connect different states with each other (i.e. the input to such a unit is a state on one map and the output an other state on an other map). Using appropriate time constants, such a network would autonomously transition between a number of states, given that it is placed in the appropriate initialization state.

There have been previous approaches to systematically embed an arbitrary DFA into neural networks. However, they have generally used unrealisitic assumptions. For example, the stability of the DFA networks of Omlin~\cite{Omlin96,Giles92} and Kremer~\cite{Kremer95} depend on the saturating non-linearity, which we have argued is not a realistic physiological requirement. Other approaches use Elman-type networks. But, these networks assume a memory layer from which activity can be artificially 'copied' at every iteration \cite{Kremer95}. Thus, the network itself has no memory. These approaches are not physiologically plausible. By contrast, our approach requires no artificial operations (like delay lines); the weights can be easily set; and their settings are independent of the DFA.

Another approach has been to project the input into a high dimensional space using randomly connected units (the liquid state machine, LSM) \cite{Maass07,Jaeger04}. Readout units that receive input from all the units in the pool are then trained to approximate the required output states. However, LSMs have fading memory and so DFAs implemented in this way can depend only on a restricted number of past inputs \cite{Natschlager02}. This deficiency can be remedied by training readout units to represent the current state, and feeding their activity back into the pool so as to maintain their current state. This circuit can implement arbitrary DFAs \cite{Maass07}. However, that approach relies on supervised learning of the weights of the output units without modification of the units in the liquid. By contrast, our network can be constructed explicitly, entirely without learning, and so allows detailed understanding of the connectivity that underlies its operation. One advantage of the liquid approach is that relatively few weights (connections to the readout) need to be changed, while all others can be held constant. This property is also true for our approach. The entire connectivity on the maps is stereotyped and does not depend on the particular state machine implemented. Adding a new state only requires connecting at least two new units reciprocally between the two sWTAs, with weights $\gamma$.

The weights of the connections in our network have no plasticity (they are static). The only changes in dynamics that can occur are thus due to activity.
Plasticity occurs on a slower timescale than changes in activity and it would thus introduce a second, slower, timescale of dynamics into the system.
Such interactions between fast (activity) and slow (plasticity) dynamics in the same networks can result in complex dynamics.
One of the questions posed by the introduction of plasticity is how multistability can be preserved (rather than convergence to a single stable state).
For our network, this remains to be explored. However, for other networks it has been shown that configurations exist which allow such multistability to
persist in the presence of plasticity  \cite{Kalitzin00}. We designed our network such that only a small number of weak connections need to be modified to
erase or introduce stable states. The remaining connectivity is homogenous and independent of function. Given multiple recurrent maps with no (or random)
coupling between them, it is thus imaginable that a simple plasticity rule can be found which learns the necessary states and the transitions between them.

Sophisticated behavior requires that a reaction to a particular stimulus 
is state dependent and so requires working memory. The frontal cortex is 
known to be crucial for this function and it is known to contain neurons 
that respond differently, according to behavioral state. One well 
studied example is the generation of memory guided sequential motor 
movements \cite{Fujii03,Shima00,Mushiake90,Barone89}. During execution 
of this task, there are two main classes of firing patterns in the supplementary motor area (SMA) \cite{Shima00}. One class of neurons fire before, during or after a particular movement is executed regardless of where in the sequence it is. Another class fire in response to a particular movement, but conditionally so: They fire only if the movement is at a particular position in the sequence, or if a particular movement was executed before. Of particular interest are two subclasses of these latter neurons. The first subclass correspond to our current state neurons, and fire for a particular location (for example, second) in the sequence. The second subclass fire between executions of movements and are conditional on the previous and the next to be executed movement \cite{Shima00}. They correspond to the transition neurons in our network. 

Neurons related to the execution of state-dependent actions have also been described in the prefrontal cortex. For example, some prefrontal units of non-human primates fire prominently only at the end of the correct execution of a complex sequence of motor actions \cite{Fujii03}. In analogy with our DFA network, these neurons could indicate that the accept state has been reached. Alternatively, they could also signal the return to the initial (start) state.

\begin{figure}
\centering
\includegraphics[angle=0,width=\linewidth]{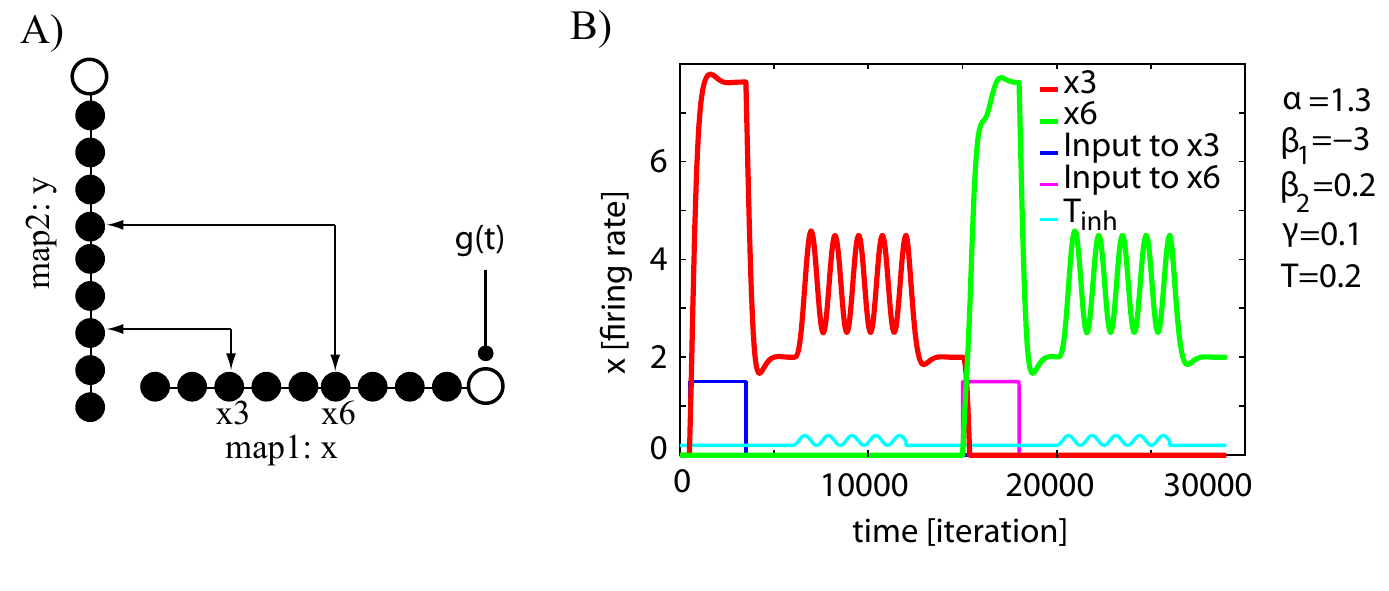}
\caption{{Demonstration of state-dependent routing. \textbf{A)} Illustration of a possible network architecture. States $x_3$ and $x_6$
represent the two possible states of the network. External inhibitory input $g(t)$ is applied to the inhibitory unit.
\textbf{B)} An analog signal applied to the inhibitory neuron ($T_{inh}$) appears only at the unit representing the current state. Depending on the current state,
the external signal appears only at the unit representing state 1 (red) or state 2 (green).
Here, the effective $T_{inh}$ is plotted, which is composed of the static non-zero firing threshold (offset from zero) as well as the external input,which varies
as a function of time. Note that the output oscillation is not generated by the network but rather is a reflection of external input.
}}
\label{figRouting}
\end{figure}

Another important property of our network is that it combines the DFA 
state property with analog sensitivity: The degree of output activation of the state neurons can be modulated by external input. This is also true during the memory state when no external input is present. Then, a positive signal $T_{inh}$ applied to the inhibitory neuron can modulate the amplitude of the output without causing the network to leave the memory state: If $T_{inh}(t)=g(t)$ is a function of time, the network output will vary according to the function $g(t)$ while the network is in the memory state. Our network thus combines processing of digital (the states) and analog (the external input) signals \cite{Hahnloser00}. One possible application of this feature is state dependent routing of arbitrary continuous signals. In this case, the signal to be routed would be supplied to the inhibitory neuron and its evoked output would be visible only at those excitatory neurons representing the current state. 
Thalamocortical input is known to project directly to inhibitory neurons \cite{Swadlow02}. Thus, the routed input signal (provided to the inhibitory neuron) would be visible only to those neurons downstream from the current state neurons. When the state changes, the analog signal will be routed to another set of downstream neurons (an example is shown in Fig \ref{figRouting}B). Modification of the inhibitory neuron firing threshold (i.e. by an additional input, see Fig \ref{figRouting}A) can also be used to dynamically enable and disable the memory state. This, for example, can be used to implement a conditional working memory. By default, $T_{inh}=0$. Only if the current stimulus is required to remain in working memory is $T_{inh}>0$.
Also note that this feature can be utilized to implement a decaying memory. By default, the persistent activity in our network does not decay as a function of time: given no external pertubations, the network remains in the memory state forever. This is clearly not desirable for some functions, such as working memory. One way to introduce working memory is to continuously decrease $T_{inh}$, relative to stimulus offset. This could, for example, be implemented
by an additional excitatory neuron (with self excitation) that feeds into the inhibitory neuron. 

Our network performs state dependent computations because it has multiple memory states. Here, we chose to implement this ability by recurrently connecting two maps. Each of the two maps has homogenous local excitatory connectivity and one global inhibitory unit. While a single map alone can have non-zero steady states (see Eq \ref{eq:steady5Main}), this state is usually not stable. Only by connecting two maps ($\gamma>0$) can stability of the memory state be guaranteed (also see appendix). Other recurrent connectivity patterns (such as multiple recurrent loops with different delays) that result in memory states are certainly possible, and remain to be explored. We chose the coupled map configuration because it requires only the replication of a simple circuit (a map) and the addition of a few (numerically small) connections between the maps and so offers a method whereby the cortex could achieve a broad range of sophisticated processing by only limited specialization of the same generic circuit.

\section{Appendix}

\subsection{Constraints on parameters}
We derive the steady state values for various configurations of the network using linear equations of the form $f(x)=x$. For simplicity, we assume that only $\alpha_1>1$ whereas all other $\alpha_i=0$ for $i>1$. The following equations describe the steady state reached after sufficient number of iterations. During the dynamic part of the systems activity, some values $x_i$ might be zero and the equations are thus not valid. The same approach as used here can, however, also be applied to this situation by replacing the rectification function with a continuously differentiable function of the form $f(x)=log(a + \exp(b (x+c)))$ ($a$,$b$,$c$ are constants). 

Also note that the approach presented here is also valid if the network contains units which have $f(x)=x$. In that case, these units are effectively non-existant and can be ignored for purposes of steady-state analysis. As long as the subset of active units $f(x)>0$ remains constant, this subset can be analyzed separately using the methods described here (i.e. piecewise analysis \cite{Hahnloser98}).

\subsection{Constraints for bounded map activity for constant input $I$ and $T=0$}

First, we define the constraints for bounded activity for constant input $I_i$ to unit $i$ only ($I_j=0$ for all $j \neq i$) and $T=0$. Steady-state implies $\dot{x_j}=0$ for all units $j=1..N$ on the map. Also $x_N>0$ (inhibitory neuron) and $x_i>0$. Solving the system of equations described by Eqs \ref{eq:recmapE},\ref{eq:recmapI} for $x_i$ results in:

\begin{equation}
x_i = \frac{I_i}{1 + \beta_1 \beta_2 - \alpha}
\label{eq:steady1}
\end{equation}

Thus, the recurrent network amplifies the input by a factor of $\frac{1}{1 + \beta_1 \beta_2 - \alpha}$ at steady state, provided $\alpha < 1 + \beta_1 \beta_2$.

\subsection{Constraints for bounded map activity for constant input $I$ and $T>0$}

Next, we define the steady-state value for constant input in the case of $T>0$. 
Solving the system of equations with this constraints for $x_i$ results in:

\begin{equation}
x_i = \frac{I_i + T(\beta_1-1)}{1 + \beta_1 \beta_2 - \alpha}
\label{eq:steady2}
\end{equation}

This equation describes the steady state value if $T>0$ for any value of $I_i$. One additionalconstraint added by this solution is that $\beta_1>1$.

Next, we describe the same two steady states for the case of two recurrently coupled maps $\mathbf{x}$ and $\mathbf{y}$ (Fig \ref{fig:recurrentCoupled}). Coupling is symmetric with weight $\gamma$. Here, we assume $x_2$ and $y_2$ are connected with $\gamma>0$. This assumption changes the dynamics of the excitatory neuron, but not the inhibitory neuron:

\begin{equation}
\tau \dot{x_i} + x_i= f( \alpha x_i + \gamma y_i + I_i - \beta_1 x_N - T)
\label{eq:recmapE2}
\end{equation}

\subsection{Constraints for bounded map activity for constant input $I$ and $T=0$}
Solving the new system of equations (Eqs \ref{eq:recmapE2} and \ref{eq:recmapI}) for $T=0$ results in:

\begin{equation}
x_i = \frac{I_i}{1 + \beta_1 \beta_2 - \alpha - \frac{\gamma^2}{1+\beta_1 \beta_2 - \alpha}}
\label{eq:steady3}
\end{equation}

Defining $K=1 + \beta_1 \beta_2 - \alpha$, the amplification factor of the recurrently coupled map is $\frac{1}{K-\frac{\gamma^2}{K}}$. Thus, the gain is well defined when $\gamma^2 < K$.

\subsection{Constraints for bounded map activity for constant input $I$ and $T>0$}
We continue to assume $K=1 + \beta_1 \beta_2 - \alpha$. Solving the same system with the constraint $T>0$ results in:

\begin{equation}
x_i = \frac{K I_i + T(\gamma+K)(\beta_1-1)}{K^2-\gamma^2}
\label{eq:steady4}
\end{equation}

This equation describes the steady state potential reached for constant input and $T>0$.  It consists of two components: the first term is the input multiplied by the gain. The second term is constant additional input that is provided by recurrent connections from other units of the network. the effect of this second term remains even after the input has been removed ($I=0$, see below).

\subsection{Existence of memory state}
A memory state exists if the excitatory neurons on both maps representing the states are non-zero ($x_i>0$, $y_i>0$), the inhibitory neurons on both maps are non-zero ($x_N>0$, $y_N>0$), and if a steady state is reached: $\dot{x}=0$, $\dot{y}=0$. Also, if the steady state is reached $x_i=y_i$ for the units $i$ representing the stable state (connected by $\gamma$). 

Setting $I=0$ (no external input) in Equation \ref{eq:steady4} results in:

\begin{equation}
x_i = \frac{T(\beta_1-1)}{1+\beta_1 \beta_2 - \alpha - \gamma}
\label{eq:steady5}
\end{equation}

This equation describes the amplitude reached by the excitatory units.  A memory state exists only if $T>0$ and $\beta_1>1$. As also seen previously, $1+\beta_1 \beta_2 > \alpha + \gamma$. Note that the absolute value of the memory state is independent of the previously applied input $I$, that is, it is entirely determined by the structure (weights) of the network.

The steady-state of the inhibitory neuron in the same situation can be calculated by Eq \ref{eq:steadyInhibit}. It adds the constraint $\alpha + \gamma > 1 + \beta_2$.

\begin{equation}
x_N = \frac{T(\alpha+\gamma-1-\beta_2)}{1+\beta_1 \beta_2 - \alpha - \gamma-1}
\label{eq:steadyInhibit}
\end{equation}

\subsection{Dynamics after constant input is removed}
\label{ref:dynamics}

The dynamics of the network following removal of the constant input, until the steady-state memory state is reached, can be characterized by the eigenvalues of the Jacobian of the entire system. Consider two recurrent maps $\mathbf{x}$ and $\mathbf{y}$ with one excitatory and one inhibitory neuron each. The system of equations shown above can be described by:

\begin{equation}
\tau \dot{\mathbf{z}}-\mathbf{z} = f( W\mathbf{z} - z + \mathbf{I} )
\label{eq:matrix1}
\end{equation}

Assume $\mathbf{z}=[x_1, y_1, x_2, y_2]$, $\tau=1$ and 

\begin{equation}
W = \left[ \begin{array}{cccc}
\alpha & \gamma & -\beta_1 & 0\\
\gamma   & \alpha & 0 & -\beta_1 \\
\beta_2  & 0        & 0 & 0 \\
0 & \beta_2 & 0 & -1
\end{array} \right]
\label{eq:W1}
\end{equation}

There are two non-zero fixed points in this simple system: the point reached with constant external input ($I>0$, Eq \ref{eq:steady4}) and the point reached after removal of the input ($I=0$, Eq \ref{eq:steady5}). First assume constant external input $I>0$ was applied for a sufficient amount of time for the system to reach steady state (Fig \ref{figTwoMaps}B). Here we are interested in the dynamics that result from the removal of this constant external input. There are two steady states the system can converge to: the memory state and the zero state (no activity). To analyze stability of the memory state, we linearize the system
at the memory state. Since the network reached its steady state with $I>0$ previously,
all $z_i>0$ and thus $f(x)=x$. Constructing the Jacobian Matrix of the system then results in

\begin{equation}
J = \left[ \begin{array}{cccc}
\alpha-1 & \gamma & -\beta_1 & 0\\
\gamma   & \alpha-1 & 0 & -\beta_1 \\
\beta_2  & 0        & -1 & 0 \\
0 & \beta_2 & 0 & -1
\end{array} \right]
\label{eq:J1}
\end{equation}

The eigenvalues $\lambda_k$ of $J$ determine which state will be reached \cite{strogatz94}. 
If the matrix is negative definite (i.e., if the real parts of all eigenvalues are $Re(\lambda_k)<0$), the memory state is asmptotically stable. Under most situations that satisfy the constraints given for the parameters we observe that this is the case. The imaginary parts $Im(\lambda_k)$ determine how the system approaches the steady state. If $Im(\lambda_k)\neq 0$ the system oscillates around the attractor with a continually reduced amplitude (because $Re(\lambda_k)<0$). The larger $max(|Im(\lambda_k)|)$ is, the larger is the initial amplitude of the oscillation. Modifying $\beta_1$ predominantly changes $Im(\lambda_k)$ while keeping $Re(\lambda_k)$ approximately constant and thus allows independent modification of the stiffness of the system. 

This approach of analyzing the dynamics is also applicable for the case of the non-linear rectification function $f(z)=max(0,z-T)$ that we use for the main part of the paper. This is because all $z_i>0$ and the jacobian matrix (\cite{strogatz94}) of
this system of equations is equal to $J$ (Eq \ref{eq:J1}) since $T$ is a constant.

\subsubsection{Assuring that the memory state is an attractor}
In this section we detail why it is necessary to have two maps ($\gamma>0$) to have a functioning memory state. From the perspective of the steady state only (Eq \ref{eq:steady5}), it is sufficient to have $T>0$ while $\gamma$ could be zero. However, the dynamics mandate that $\gamma>0$.

The four eigenvalues of the Jacobian matrix $J$ (Eq \ref{eq:J1}) are described by
\begin{equation}
\lambda_i = -1 \pm \frac{1}{2} \gamma + \frac{1}{2} \alpha \pm \frac{1}{2} \sqrt{\gamma^2 \pm 2 \gamma + \alpha + \alpha^2 - 4\beta_2\beta_1}
\label{eq:eigvals1}
\end{equation}

Note that two of the $\pm$ are in front of terms which disappear in the case of $\gamma=0$. This results in only two unique eigenvalues and the node is thus degenerate (and unstable). For the node to be stable four unique eigenvalues (with the properties described above) are required and thus it is necessary that $\gamma>0$.

The trace of $J$ (the sum of all eigenvalues $\lambda_i$) equals $2\alpha-4$ and is independent of $\gamma$. It is required that the trace is smaller than 0 (for a state to be stable) and thus $\alpha<2$.

\subsection{Constraints for bounded map activity for permanently active transition neurons, $I=0$ and $T>0$}
In the presence of an active transition neuron (TN), the effective recurrent connectivity between the two maps is increased.
In this section we show the constraints on the parameters of the TN such that the map activity remains bounded.
In the practice, the external input to the TN is only active for a short time. For the purposes of deriving boundaries that guarantee
stability, however, we assume a permanently active TN (as a worst case scenario). Also, we assume that the TN receives input from and projects
to map neurons that represent a state that is currently active. This is relevant for state transformations of symbols that project to the same state
(loops). 
We use the minimal system, as described in the previous section \ref{ref:dynamics}. It consists of two recurrent maps with one inhibitory ($y_1, y_2$) and excitatory unit each ($x_1, x_2$).The two excitatory units on both maps are recurrently connected with weights $\gamma$.  
Here, we add one unit to the system: the TN $p_1$. It receives input from the excitatory unit of one map $x_2$ with weight $\phi$ and projects
to the excitatory unit of the other map $x_1$ with the same weight (see Eq \ref{eq:pointers} and Fig \ref{fig:recurrentTNs}). 
Assuming that the threshold $T_p$ equals the amplitude of the external input (as we assume throughout), a permanently active TN has effectively $T_p=0$ (since input is present). 
The steady-state amplitude of the excitatory map unit $x_1$ and $x_2$ is then described by:

\begin{equation}
x_i = \frac{T(\beta_1-1) (K+A)}{K^2-A\gamma^2}
\label{eq:steadyTN}
\end{equation}

with $A=\phi^2+\gamma$ and K as defined previously. In steady state, $\dot{x_i}=0$, $\dot{y_i}=0$, $\dot{p_1}=0$ while $x_i>0,y_i>0,p_1>0$.
Solving for $\phi$ shows that the following needs to hold:

\begin{equation}
\phi < \sqrt{\frac{K^2-\gamma^2}{\gamma}}
\label{eq:steadyTN2}
\end{equation}

Also, $\phi>0$. For example, for the numerical values of the weights used in Fig \ref{figStateSwitch}, $0<\phi<0.8944$.

\subsection{Speed of convergence}
The network requires time to reach steady state. The time (and thus the number of iterations of numerical integration) it takes for the network to converge is important because it determines the minimum time an external input (either to the transition neuron or to the state neuron) needs to be applied to assure a state change. 

The rate limiting step for convergence is the dimension with the smallest absolute eigenvalue, i.e. $\min_i |\lambda_i|$.
The smaller this value, the longer the system will take to converge. The eigenvalues are a function of the weights (Eq \ref{eq:eigvals1}) and the setting
of the weights thus has an influence on the speed of convergence. Note that the size of the network is not a factor, since the number of active units is
always the same, regardless of which state is currently active. 

We confirmed numerically how fast the network converges to the application of external input and state switches.
We tested random DFAs with 2-40 states and found that the time required for the network to converge to the memory state
does not depend on the number of states nor the distance (on the map) between the states. 
Euler integration with $\delta=0.05$ revealed a relaxation time of the following number of iterations: 143 ($7.15\tau$) after initialization, 305 ($15.2\tau$) after a state switch and 351 ($17.6\tau$) after a loop (state transformation that points to itself). These numbers remain the same irrespective of the number of states and transitions represented by the network. The parameters of the network were as shown in Fig \ref{figStateSwitch}.

\subsection{Summary of constraints}

Two recurrently coupled maps with coupling weight $\gamma$, inhibitory loop constants $\beta_1$, $\beta_2$ and total excitatory input from the same map $\alpha$ have bounded activity and a stable memory if the following conditions are met:

\begin{equation}
\gamma < 1+\beta_1 \beta_2 - \alpha
\label{eq:cond1}
\end{equation}

\begin{equation}
\beta_1 > 1
\label{eq:cond3}
\end{equation}

\begin{equation}
T > 0
\label{eq:cond4}
\end{equation}

\begin{equation}
\gamma > 0
\label{eq:cond5}
\end{equation}

\begin{equation}
\alpha < 2
\label{eq:cond6}
\end{equation}

\section{Acknowledgements} 
We thank the participants of the Neuromorphic Engineering workshop in Cappo Caccia, Sardinia (CNE 2008) for discussion and Robert Rohrkemper
for help with figure preparation. This research was funded by the European Union (DAISY project, FP6-2005-015803) 
and the California Institute of Technology.

\end{document}